\documentclass[final,5p,times,twocolumn]{elsarticle}

\journal{Nuclear Instruments and Methods A}

\usepackage{graphicx}

\usepackage{hyperref}

\begin{document}

\begin{frontmatter}

\title{Studies on a silicon-photomultiplier-based camera for Imaging Atmospheric Cherenkov Telescopes}

\author[A,B]{C.~Arcaro*$^,$}
\author[B]{D.~Corti}
\author[B,C,D]{A.~De Angelis}
\author[A,B]{M.~Doro}
\author[E]{C.~Manea}
\author[A,B]{M.~Mariotti}
\author[A,B]{R.~Rando}
\author[B]{I.~Reichardt}
\author[B]{D.~Tescaro}

\address[A]{Universit\`a di Padova, I-35131 Padova, Italy}
\address[B]{INFN, I-35131 Padova, Italy}
\address[C]{INAF Padova, Italy}
\address[D]{LIP/IST Lisboa, Portugal}
\address[E]{INFN TIFPA, I-38123 Povo, Italy}

\begin{abstract}
Imaging Atmospheric Cherenkov Telescopes (IACTs) represent a class of instruments which are dedicated to the ground-based observation of cosmic VHE gamma ray emission based on the detection of the Cherenkov radiation produced in the interaction of gamma rays with the Earth atmosphere. One of the key elements of such instruments is a pixelized focal-plane camera consisting of photodetectors. To date, photomultiplier tubes (PMTs) have been the common choice given their high photon detection efficiency (PDE) and fast time response. Recently, silicon photomultipliers (SiPMs) are emerging as an alternative. This rapidly evolving technology has strong potential to become superior to that based on PMTs in terms of PDE, which would further improve the sensitivity of IACTs, and see a price reduction per square millimeter of detector area. We are working to develop a SiPM-based module for the focal-plane cameras of the MAGIC telescopes to probe this technology for IACTs with large focal plane cameras of an area of few square meters. We will describe the solutions we are exploring in order to balance a competitive performance with a minimal impact on the overall MAGIC camera design using ray tracing simulations. We further present a comparative study of the overall light throughput based on Monte Carlo simulations and considering the properties of the major hardware elements of an IACT.
\end{abstract}


\end{frontmatter}

\section{Introduction}
\makeatletter{\renewcommand*{\@makefnmark}{}
\footnotetext{* Corresponding author at: Universit\`a di Padova, I-35131 Padova, Italy; INFN, I-35131 Padova, Italy; \href{mailto:cornelia.arcaro@pd.infn.it}{cornelia.arcaro@pd.infn.it}}\makeatother}
In recent years ground-based very high energy (VHE, $E\ge100$\,GeV) gamma-ray astronomy has experienced a major breakthrough, as demonstrated by the impressive astrophysical results obtained with Imaging Atmospheric Cherenkov Telescopes (IACTs) like H.E.S.S., MAGIC or VERITAS~\citep{aharonian08}. IACTs, with a large single mirror dish of the order of few tens of meters in diameter and a camera area of the order of square meters, as MAGIC, are suited to cover the lower end of the VHE energy range. They are designed to collect the maximum number of Cherenkov photons possible by means of a large mirror dish of high reflectance in the wavelength range where Cherenkov light is emitted. Such light is generated by the interaction of VHE gamma rays with the Earth atmosphere and emitted as light flashes of few nanoseconds. To maximize the benefit of the large collecting area, the efficiency of the photodetectors has thus to be as high as possible in the wavelength range of interest. The baseline design of a large IACT includes commonly a focal-plane camera based on photomultiplier tubes (PMTs) given its photon detection efficiency (PDE) and fast time response. Each pixel usually incorporates a PMT, the corresponding readout electronics and a light concentrator, usually a Winston cone made of material with high reflectivity in the wavelength range of Cherenkov radiation, to guide the light reflected from the mirror dish to the effective area of the photodetector and to reduce the acceptance of stray light. 

Silicon photomultipliers (SiPMs), a rapidly evolving class of solid-state photon sensors, are very promising detectors for Cherenkov applications, thanks to high PDE and high gain, low operating voltage, tolerance to high illumination levels, good single-photon sensitivity, fast response, and low-amplitude afterpulses. Drawbacks of such photosensors are a higher capacitance and higher cross talk.

The feasibility of SiPMs for IACT applications is being explored for telescopes with a small focal plane camera, i.e., with an area of the order of few hundreds of square centimeters and a pixel size of few millimeters, and telescopes with double-mirror Schwarzschild-Couder layout~\citep{catalano13}. The First G-APD Cherenkov Telescope (FACT;~\citep{bretz14}) has shown that SiPMs show a stable and reliable long-term performance~\citep{biland16}. However, large size SiPM-based IACT cameras, with an area of few square meters and a pixel size of the order of several centimeters, have not yet been built. The problems encountered in the construction of large, monolithic photon-sensitive silicon devices are mainly related to the large detector capacity, which significantly increases the noise level of the sensor, and to the low device yield due to the natural occurrence of defects on the silicon wafers.  A good compromise is therefore to segment the detector into elements of small size, and summing the output signal electronically, e.g., through an analog high speed adder.

The goal of our research is to develop a SiPM-based module for the focal-plane camera of large-dish IACTs. We thus developed a first prototype for the MAGIC camera based on the concept to balance a competitive performance with a minimal impact on the overall camera design. 

One additional issue that needs to be addressed is the SiPM sensitivity to red photons: for IACT applications, sensitivity
beyond the range of the Cherenkov spectrum largely increases the background due to the abundance
of atmospheric photons. We thus compare the performance among the PMT and SiPM-based pixel design using ray tracing simulations and study the overall performance of the IACT, including a discussion on the effect of the Night Sky Background (NSB).
\begin{figure}[t!]
\resizebox{\hsize}{!}{\includegraphics{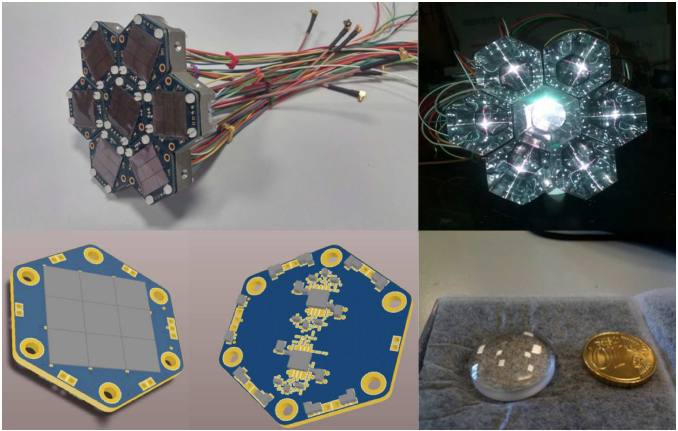}}
\caption{SiPM-based prototype pixel cluster (\textit{top left}) designed for the MAGIC focal-plane camera with seven pixels each consisting of  nine $6\times6$\,mm$^2$ sensors, which are glued onto the PCB (\textit{bottom left}), and equipped with the MAGIC Winston cone (\textit{top right}).  A fused silica PCL (\textit{bottom right}) is placed on all pixels except the central one.}
\label{fig:1}
\end{figure}
\section{SiPM-based module}\label{sec:2}

The approach described here involves covering the active area of the standard MAGIC pixel with several SiPM sensors, and summing their individual analog signals into one single output, to be processed by the existing PMT readout chain. Each segmented sensor covers 0.1$^\circ$   and matches the exit window of the existing light concentrator with hexagonal entrance pupil, with each SiPM segment covering a total active area of few square centimeters.  

A first cluster was designed and assembled at INFN Padua. The design is based on nine SiPM sensors from FBK (Fondazione Bruno Kessler\footnote{\href{http://www.fbk.eu/}{http://www.fbk.eu/}})
of $6\times6$\,mm$^2$ size for each channel (see Fig.~\ref{fig:1}).  

\begin{figure}[t!]
\resizebox{\hsize}{!}{\includegraphics{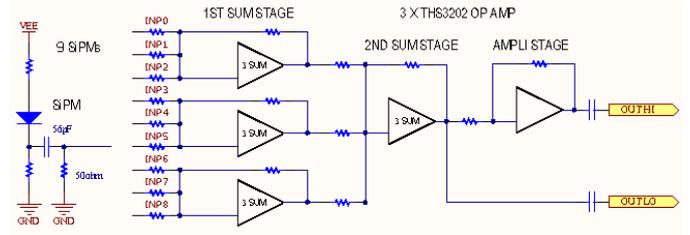}}
\caption{Adder circuit based on a two stage operational amplifier design. In the current prototype, the signals
from 9 $6\times6$\,mm$^2$ SiPMs are filtered (high-pass) and summed in groups of three into the first sum stage; 3 of such first-stage
sums are then summed into one final output in the second-stage adder. A further amplification stage is possible to split the output to have two gain levels.}
\label{fig:2}
\end{figure}

The adder circuit, which is realized on a printed circuit board (PCB), is optimized to keep the shape of the individual SiPMs signals narrow, which is important for signal-to-noise ratio, and for a low power consumption. A first sum stage based on a fast, low-noise operational amplifier collects and sums the signals from three SiPM sensors, with some gain. A second stage collects and sums the signals from the first sum signals, to give a single output of the nine sensors; some additional gain is applied here. Finally, one last additional gain stage can be used to implement a two-gain output design (Fig.~\ref{fig:2}). The power consumption for the 9-channel version is $\sim$400\,mW, SiPM power excluded, while the pulse width is $\sim$4\,ns.

\begin{figure*}[t!]
\center
\resizebox{\hsize}{!}{\includegraphics{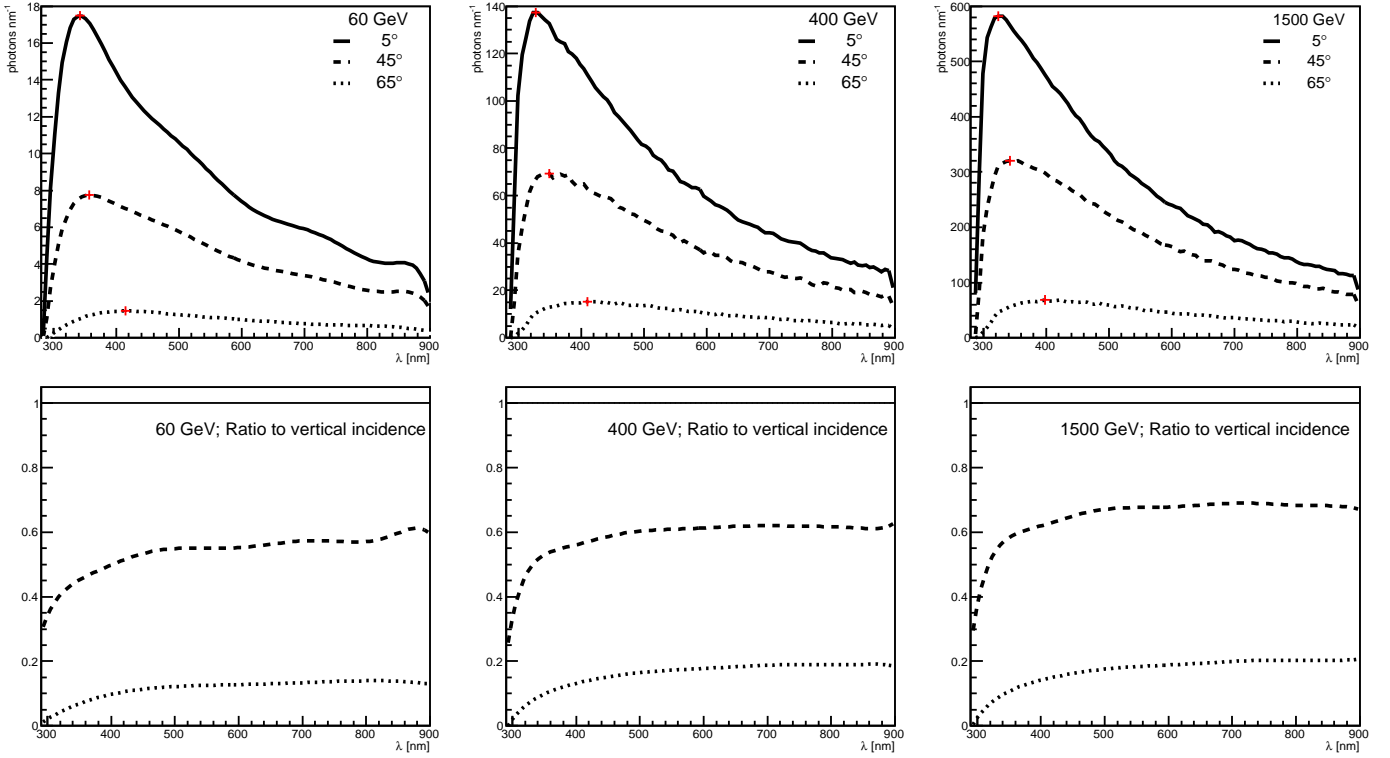}}
\caption{Cherenkov spectrum at ground in function of the energy and direction of the primary gamma. \textit{Top panels:} the
      total number of photons in a standard telescope camera for three
      different energies: 60\,GeV (\textit{left}), 400\,GeV (\textit{middle}) and 1500\,GeV
      (\textit{right}) and for three zenith angles (5, 45 and 65
     $^\circ$; solid, dashed and dotted line, respectively). The
      markers (red) indicate the spectral peaks. \textit{Bottom panels:} the
      total number of photons of the individual spectra is normalized with respect to the 5$^\circ$ case.}
\label{fig:3}
\end{figure*}
The goal of the camera optics is to map the focal plane into the pixelized camera avoiding dead areas between pixels and to compress the hexagonal entrance pupil area into the sensor area. The area compression ratio is limited by the optical input acceptance and the angular acceptance of SiPMs. To optimize the performance of the optical elements of the focal-plane camera pixels in terms of efficiency in the area compression ratio, detailed ray tracing simulation have been performed and different possible solutions are under investigation for the MAGIC telescopes. 

The design of the Winston cones is usually optimized to compress the entrance area to the surface of the PMTs, which is usually curved (typically hemispheric). SiPM sensors are usually flat and their PDE decreases with increasing incident angle. As the compression from the entrance of the Winston cone to its exit aperture results in photons hitting the sensor surface with large incident angle, these Winston cones are not optimal for this type of photosensors.

A simple solution to adapt the Winston cone, which has been explored for the realization of the MAGIC SiPM-based prototype pixel, is to add a fused silica plane-convex lens (PCL) on top of the silicon sensor, creating a curved surface similar to the entrance window of PMTs (Fig.~\ref{fig:1}). Ray tracing simulations using the TracePro software provided by the Lambda Research Corporation were used to determine the profile of the lens that maximizes the efficiency. 

Other solutions consider alternative Winston cone designs as well as solid concentrators of adequate material, that is lightweight and transparent in the near ultraviolet (UV), with layouts adapted to the possible dimension of a SiPM-based pixel following~\cite{welford1989} through ray tracing simulations and are currently under investigation. The simulations are steered considering the number of rays hitting the sensor, the area to which they are compressed by the optics under testing and their incident angle distribution. Among other elements considered in such simulations, we define a Cherenkov-light emitting light source and plug in the properties of the material considered for the optical element under study and the sensor properties, that is its PDE and angular response. The spectral distribution of those rays hitting the sensor surface are then weighed by these properties. The weighed ray spectrum is compared to that from simulations performed with MAGIC standard pixel settings by using the integral of those spectra as a proxy for the expected pixel performance.

Our design study is somehow constrained by the existing pixel layout, that is by its entrance aperture and the angular acceptance, being $\sim$30\, mm and 40$^\circ$, respectively, wherefore our simulations foresee an insufficient light compression in the case of a classical Winston cone and too high absorption in the ultraviolet band in the case of solid optics due to its length of several tens of millimeters. Therefore, a combination of a Winston cone with solid optics is currently investigated.

The first cluster prototype, for which slow control, optical transmitters, design of mechanics and HV supply were provided by the MAGIC MPI-Munich team, was installed in the MAGIC camera on La Palma next to the PMT clusters in October 2015, with the Winston cone being adapted by a PCL as previously described (see Sec.~\ref{sec:2}), to test the performance of the prototype and to validate the pixel layout in a functional IACT. As the output signals of the SiPM cluster were adjusted to be similar to those from the PMT cluster, the obtained data, such as the pixel temperature and the charge, are recorded by the regular MAGIC readout, allowing performance and long term tests to compare in detail the PMT and SiPM clusters, which will be crucial for the decision on the next steps in SiPM cluster developments. The analysis of these data is currently ongoing.

\section{Performance studies}

Since the total throughput on the pixels of the focal-plane camera of an IACT depends both on the mirror dish reflectivity and on the PDE of the light sensor, we investigated the interplay between these two hardware elements, for different energies and arrival direction of the primary gamma ray including the Cherenkov light extinction through the atmosphere (Fig.~\ref{fig:3}). 

\begin{figure}[h!]
\resizebox{\hsize}{!}{\includegraphics{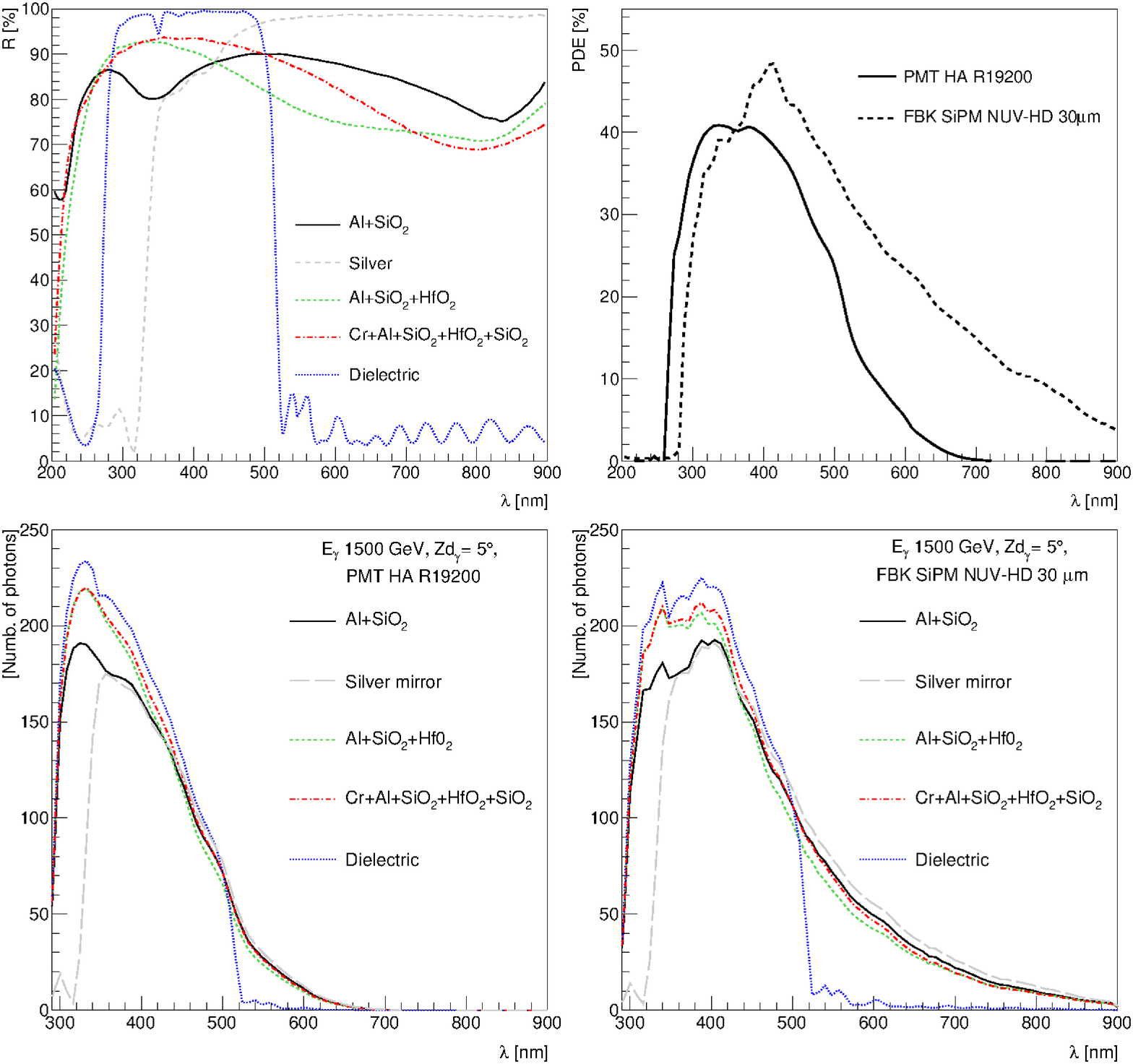}}
\caption{\textit{Top left:} Reflectivity curves considered in our study. The Al+SiO$_2$ (black solid line) is taken
      from~\cite{bastieri2005} and is considered as reference. The silver coating (long dashed gray) is taken from
      MAGIC technological tests and reported in~\cite{barrio1998}. The red dot-dashed and green short-dashed lines represent two metallic coatings (Al+SiO$_2$+HfO$_2$ and Cr+Al+SiO$_2$+HfO$_2$+SiO$_2$, respectively), while the doted blue indicates a dielectric coating. All three coatings are under test for the Cherenkov Telescope Array (CTA) project and are described in~\cite{foerster2013}. \textit{Top right:} PDE  for a reference PMT and SiPM for this paper. The PMT is a Hamamatsu 19200 (from~\cite{okumura2015}) planned for the use in the large-size telescope of the CTA project. The SiPM (from~\cite{otte2016}) is a 30 $\mu$m coated FBK sensor, currently under development and also candidate for use in CTA. \textit{Bottom:} Fraction of photoelectrons that survive the reflection onto the primary mirror of the telescope and the photo-conversion into the PMT (\textit{left}) and SiPM (\textit{right}) for a primary gamma of 1500 GeV and 5$^\circ$ zenith angle direction.}
\label{fig:4}
\end{figure}

\begin{figure}[h!]
\resizebox{\hsize}{!}{\includegraphics{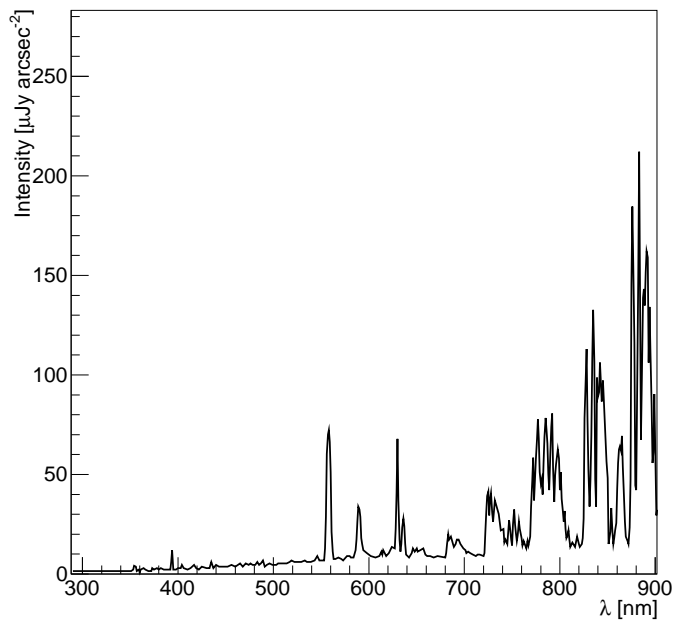}}
\caption{The spectrum of the NSB at La Palma from~\cite{benn1998}.}
\label{fig:5}
\end{figure}

As working example, we compared possible different reflective surfaces for future IACTs (Fig.~\ref{fig:4}). We repeated the exercise considering the PDE of PMTs and SiPMs and compared the total amount of Cherenkov photons detected. For simplicity, we did not consider the performance of the light guide in these studies and neglected the cross talk. In the same manner, we studied the importance of the NSB (Fig.~\ref{fig:5}). 

\begin{figure}[h!]
\resizebox{\hsize}{!}{\includegraphics{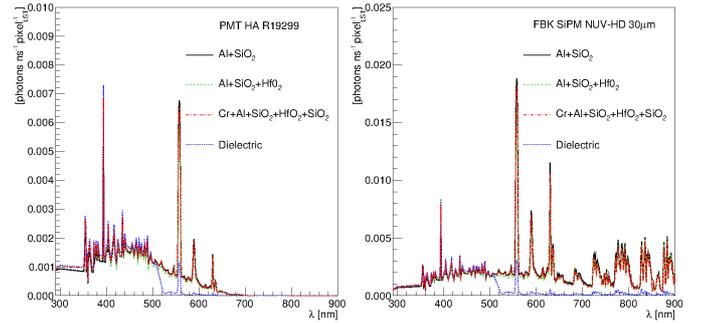}}
\caption{Rate of the NSB per nanosecond per a camera pixel of the large-size telescope (LST) of the Cherenkov Telescope Array project weighed with individual mirror
      reflectivity curves and the PDE of a PMT
      (\textit{left}) and a SiPM (\textit{right}).}
\label{fig:6}
\end{figure}

We concluded that the choice of the current generation of SiPMs over PMTs would lead to an enhanced throughput irrespective of the mirror coating, due to the enhanced PDE of SiPMs in a wavelength range of the Cherenkov spectrum ($\sim$290 - 900\,nm), in particular at wavelengths larger than 700 nm where the PDE of and PMT drops to zero.

This leads consequently to an enhanced detection of the NSB whose dominant emission occurs in the wavelength range from $\sim$720 - 1000\,nm (Fig.~\ref{fig:6}), except for the dielectric coating, which suppresses the NSB contribution at the cost of a decreased Cherenkov light detection. 

This may have an effect on the energy threshold of the telescope in presence of large NSB, due to the small signal-to-noise ratio at the lowest energies. To quantify properly the effect of the NSB on the detector sensitivity, we are currently performing Monte Carlo simulations. 

\section*{Acknowledgements}
We thankfully acknowledge the cooperation of MAGIC experiment in the installation and test of the SiPM cluster prototype in their telescope.
In particular we acknowledge the help of the MAGIC MPI-Munich team in the design of the cluster module.\\
Funds for this study have been provided within the Progetto Premiale TECHE.it (TElescopi CHErenkov made in Italy) by the Italian Ministry of Education, University and Research.

\end{document}